\documentclass[11pt,twocolumn,twoside]{template} 
\templatetype{twocolumns} 
\title{Simulating Social Media Using Large Language Models to Evaluate Alternative News Feed Algorithms}

\usepackage{longtable}

\author[1,c]{Petter Törnberg}
\author[2]{Diliara Valeeva}
\author[2]{Justus Uitermark}
\author[3]{Christopher Bail}
\affil[1]{University of Amsterdam, Institute of Language, Logic and Computation (ILLC). \textsuperscript{c}p.tornberg@uva.nl. ILLC. P.O. Box 94242, 1090 GE Amsterdam}
\affil[2]{University of Amsterdam, Amsterdam Institute for Social Science Research (AISSR)}
\affil[3]{Duke University, Department of Sociology, Political Science, and Public Policy.}

\leadauthor{Petter Törnberg} 


\begin{abstract}
\textbf{Abstract.} Social media is often criticized for amplifying toxic discourse and discouraging constructive conversations. But designing social media platforms to promote better conversations is inherently challenging. This paper asks whether simulating social media through a combination of Large Language Models (LLM) and Agent-Based Modeling can help researchers study how different news feed algorithms shape the quality of online conversations. We create realistic personas using data from the American National Election Study to populate simulated social media platforms. Next, we prompt the agents to read and share news articles --- and like or comment upon each other's messages --- within three platforms that use different news feed algorithms. In the first platform, users see the most liked and commented posts from users whom they follow. In the second, they see posts from all users --- even those outside their own network. The third platform employs a novel ``bridging'' algorithm that highlights posts that are liked by people with opposing political views. We find this bridging algorithm promotes more constructive, non-toxic, conversation across political divides than the other two models. Though further research is needed to evaluate these findings, we argue that LLMs hold considerable potential to improve simulation research on social media and many other complex social settings.
\end{abstract}

\begin{document}
\maketitle 
\thispagestyle{firststyle}
\ifthenelse{\boolean{shortarticle}}{\ifthenelse{\boolean{singlecolumn}}{\abscontentformatted}{\abscontent}}{}
\dropcap{S}ocial media platforms are routinely blamed for reducing the capacity of citizens to engage in productive conversations \citep{bail_exposure_2018,finkel_political_2020,lorenz_spreen_systematic_2022,persily_social_2020, settle_frenemies_2018}. Many argue that Facebook, Twitter, and other platforms trap people within ``echo chambers'' or ``filter bubbles'' that prevent exposure to opposing views \citep{barbera_birds_2015,cinelli_echo_2021,sunstein_republic_2002}. Algorithms are similarly blamed for incentivizing toxic, combative conversations by highlighting posts that provoke engagement, even if such engagement is largely negative or only involves people ``preaching to the choir'' \citep{brady_emotion_2017,rathje_out_group_2021, shmargad_sorting_2020, tornberg_how_2022}. Such criticism has decreased public trust in social media platforms, inspired policy makers to consider regulating them, and provoked considerable debate within the industry itself \citep{franks_beyond_2021,masullo_what_2022,persily_social_2020}.

Although causal evidence of the negative impacts of social media algorithms on public conversation is limited \citep{eady_how_2019,gonzalez_bailon_asymmetric_2023,guess_how_2023,guess_reshares_2023,haidt_social_2022,lorenz_spreen_systematic_2022,nyhan_like_minded_2023}, few scholars would disagree that such algorithms could be redesigned to promote more constructive dialogue \citep{bail_breaking_2021,dorr_incentivizing_2023,masullo_what_2022}. Early observers of social media argued these platforms would improve democracy by enabling people to connect across social divides \citep{benkler_wealth_2007,castells_networks_2012}. Yet most social media companies were not created to support such high-minded goals \citep{aral_hype_2020,bail_breaking_2021}. For example, Facebook famously evolved from a platform designed to help college students rate each other's physical attractiveness while Twitter was created to help friends share SMS messages with each other in a more efficient manner. None of the world's largest platforms --- including TikTok, Instagram, and Youtube --- were designed to promote a constructive public sphere. Assessing the impact of these platforms on public conversation may therefore be less productive than exploring new alternatives.

In this article, we study how social media could be redesigned to promote more constructive conversations. Specifically, we ask: can social media platforms increase interaction between people with opposing views without increasing toxic or uncivil behavior? Deliberation and compromise among people with diverse viewpoints is a cornerstone of democratic theory \citep{dewey_public_1927,habermas_structural_1989}. Unfortunately, some studies indicate exposing people to opposing views can increase toxicity, instead of constructive conversation \citep{bail_exposure_2018,yang_mitigating_2020,di_tella_does_2021}. We propose that algorithms that reward consensus --- that is, likes from users who have different viewpoints --- can enable cross-partisan interaction without resulting in toxic and conflictual discourse.

Examining the impact of algorithms on online conversations presents numerous challenges. Social media platforms are complex systems with emergent properties that result from the interaction of large groups of people. Though observational data can be useful for mapping emergent processes, they are of limited utility for identifying their underlying causes of group behaviors --- let alone assessing alternative social structures such as not yet existing social media algorithms \citep{coleman_stewardship_2021,byrne_complexity_2023}. Large-scale randomized controlled trials can shed light on the impact of algorithms on individual social media users \citep{gonzalez_bailon_asymmetric_2023,guess_reshares_2023}, but such interventions cannot examine how algorithms affect the many interactions among users from which public discourse emerges \citep{coleman_stewardship_2021,guess_how_2023,garcia_influence_2023}. Large-scale experiments are further limited by the corporate interests of social media platforms and the risks such experiments may pose to human subjects \citep{wagner_independence_2023}. Finally, the vast majority of researchers do not have access to social media companies, and the amount of data that is publicly shared continues to decrease precipitously \citep{freelon_computational_2018}.

To address these challenges, we ask whether recent advances in artificial intelligence can allow researchers to conduct simulated experiments on social media platforms. Such simulations employ Agent-Based Models (ABM), tools that for many years have been used to create synthetic social environments where simulated human agents interact with each other \citep{tornberg_how_2022,epstein_generative_2006,macy_factors_2002}. Researchers have already employed ABMs to capture emergent properties of social media --- such as the tendency of intra-group bonding to create polarization \citep{axelrod_dissemination_1997,noorazar_classical_2020}. Yet such simulations do not capture human discourse \citep{byrne_complexity_2023,goldspink_social_2007}. Agents within ABMs usually follow simple rules, and cannot reason, interpret social contexts, or use language --- much less engage in conversations. This severely limits the capacity of these models to simulate realistic public discourse \citep{conte_agent_based_2014}.

The advent of Large Language Models (LLMs) may provide potent opportunities to create more realistic simulations of public discourse \citep{bail_can_2023,grossmann_ai_2023,park_generative_2023}. LLMs are a new form of generative artificial intelligence, trained on large amounts of human language to realistically simulate human conversations. Studies have found that state-of-the-art LLMs demonstrate several emergent capacities \citep{wei_emergent_2022}. For example, LLMs can generate persuasive arguments \cite{palmer_large_2023}, draw on contextual knowledge \citep{tornberg_chatgpt_4_2023}, perform basic reasoning tasks \citep{bubeck_sparks_2023}, and perhaps even simulate survey responses \citep{argyle_out_2023}. Such capacities may thus address many of the aforementioned limitations of ABMs for the study of social media. In this study, we ask whether a large group of LLMs that interact with each other can be used to simulate a social media platform, and compare how different news feed algorithms shape the quality of conversations therein.

We create three synthetic social media environments designed to emulate text-based platforms such as Twitter, Threads, Mastodon, or Bluesky for approximately one day during July 2020. Each environment is shaped by a different algorithm that determines which posts are shown to users within their news feed --- and we create LLM-based agents that post, like, and comment within each environment. In the first synthetic platform, users see the most liked and commented posts from users whom they follow. The second shows all high-engagement posts--- even those that are not generated by accounts the user follows. The third platform employs a novel ``bridging'' algorithm that highlights posts liked across partisan lines. To calibrate the model to the US electorate, we construct a persona for each agent using data from the 2020 American National Election Study. These include information about the demographic characteristics of those who use social media as well as their political beliefs and cultural preferences. We leverage the capacity of LLMs to simulate survey responses to assign each agent a fictitious name, non-political interests (e.g. sports, television shows, and hobbies), and personality traits.

Our goal is not to recreate real social media conversations by prompting LLMs to simulate real human users, but to simulate a range of possible outcomes of alternative news feed designs that could guide future research in this area. We use a combination of well-established text analysis tools and behavioral measures to assess the toxicity and amount of cross-party dialogue within each synthetic environment --- and describe our plans to further validate our approach with real human coders. Though further research is needed, we find tentative evidence that conventional forms of social media either prevent cross-party interaction or increase toxicity. The bridging algorithm, by contrast, both increases cross-party engagement among our simulated agents and reduces the toxicity of the language they use when interacting with each other. These results may have important implications for the study of polarization, political communication, and social media as well as the broader field of computational social science.

\section{Can Bridging Algorithms Improve Conversations on Social Media?}
Most social media platforms provide users with some degree of control over who they will receive updates from within their news feed by ``following'' or ``friending'' people. Though this allows users to customize their feeds to produce information that is entertaining or useful to them, many worry this may limit users' exposure to those with alternative views \citep{barbera_birds_2015,cinelli_echo_2021,sunstein_republic_2002}. Though there is mixed evidence for this claim \citep{eady_how_2019}, the idea that interaction with those who do not share one's views has become a cornerstone of democratic theory. Or, to paraphrase John Stuart Mill, conversation can be described as the ``soul of democracy.'' A large literature indicates intergroup contact can reduce prejudice between members of rival groups \citep{imperato_allport_2021}. Most scholars thus believe that the exchange of arguments among people with different views is a necessary component of constructive public conversations.

Some public figures have therefore called upon social media platforms to encourage users to break their echo chambers. But some studies indicate that increasing exposure to opposing views on social media might actually increase intergroup tensions \citep{bail_exposure_2018,di_tella_does_2021,yang_mitigating_2020,di_tella_does_2021}. Rather than encouraging people to calmly reflect upon the merits of those with opposing views, such exposure may increase the likelihood that users attack their rivals --- a phenomenon that could be compounded by news feed algorithms that reward engagement, even if it is negative or conflictual in nature \citep{brady_how_2021,huszar_algorithmic_2022,rathje_out_group_2021,tornberg_how_2022}. Put differently, encouraging cross-partisan interaction among social media users may produce more toxic discourse, which we define here as conversations that are rude, disrespectful, or unreasonable.

Using social media algorithms to foster constructive public conversations may therefore require navigating the Scylla and Charybdis of partisan isolation and toxic discourse. Scholars have recently proposed that the algorithms guiding social media news feeds could be redesigned in order to promote posts that receive positive engagement from people with different views and backgrounds \citep{bail_breaking_2021,ovadya_bridging_2023,wojcik2022birdwatch}. Such ``bridging'' algorithms could, for example, amplify posts that receive positive reactions from across the political spectrum. The theory holds that bridging algorithms would decrease the capacity of political extremists to hijack social media conversations by provoking others. Such algorithms may also incentivize more moderate users --- who are often intimidated or overshadowed by extremists --- to express their views more frequently or enthusiastically by providing them with a signal that their messages are appreciated by others.

In short, it is challenging to nurture constructive conversations online as there seems to be a trade-off between maximizing the range of political viewpoints and minimizing the toxicity of conversations. Is it possible to resolve this trade-off? To answer this question, we develop and test a bridging algorithm designed to promote inter-partisan exposure \emph{and} minimize toxicity.

\section{Simulating Social Media Platforms Using Large Language Models}
Agent-Based Models (ABM) have been used to study complex social systems for decades, enabling the identification of emergent, group-level mechanisms that could not be predicted using the characteristics of individuals within a social system alone. That is, ABMs can simulate how individual-level behaviors affect each other, creating large-scale social patterns that are difficult to anticipate without simulating interactions at scale \citep{conte_agent_based_2014,cioffi_revilla_bigger_2016}. While ABMs cannot predict precise outcomes or fully capture social reality, this is not their intended use. Instead, ABMs allow researchers to identify social mechanisms by observing how local interactions lead to global patterns. This bottom-up approach offers a granular insight into the causal pathways and feedback loops that drive system behavior \citep{keuschnigg_analytical_2018}. In this way, ABMs can be understood as enabling a kind of ``model assisted design,'' where researchers identify a range of possible outcomes that can inform more deductive experimental research. However, a common criticism of ABMs is that they often rely on simplified assumptions about individual behavior and decision-making processes, potentially leading to oversimplified or unrealistic representations of complex social phenomena and interactions. Agents have lacked the capacity to reason, discuss, and engage in realistic forms of social interaction.

Large Language Models are a novel form of generative artificial intelligence, built on deep neural networks and trained to predict the next word in a sequence given a set of preceding words. LLMs learn such associations by ingesting large amounts of text --- typically from the internet. As a result, LLMs are capable of engaging in realistic conversation, suggesting a potential complementarity with ABMs, in particular when it comes to simulating the dynamics of public discourse.

We draw on this complementarity to create synthetic platforms designed to mimic social media users in the United States on a single day in July 2020 (see Figure \ref{fig:overview} for an overview; see Appendix for a detailed description of our methodology). We chose this context because the majority of research on social media examines the United States \citep{lorenz_spreen_systematic_2022}, and previous studies indicate large language models excel at impersonating Americans \citep{tornberg_chatgpt_4_2023}. To provide a high fidelity representation of this population, we employed data from the American National Election Study (ANES) --- a large, nationally representative study of political attitudes and behaviors that includes measures of social media usage as well as demographic variables.

\begin{figure*}
\centering
\includegraphics[width=0.5\textwidth,angle=90,trim=4cm 3cm 3cm 4cm]{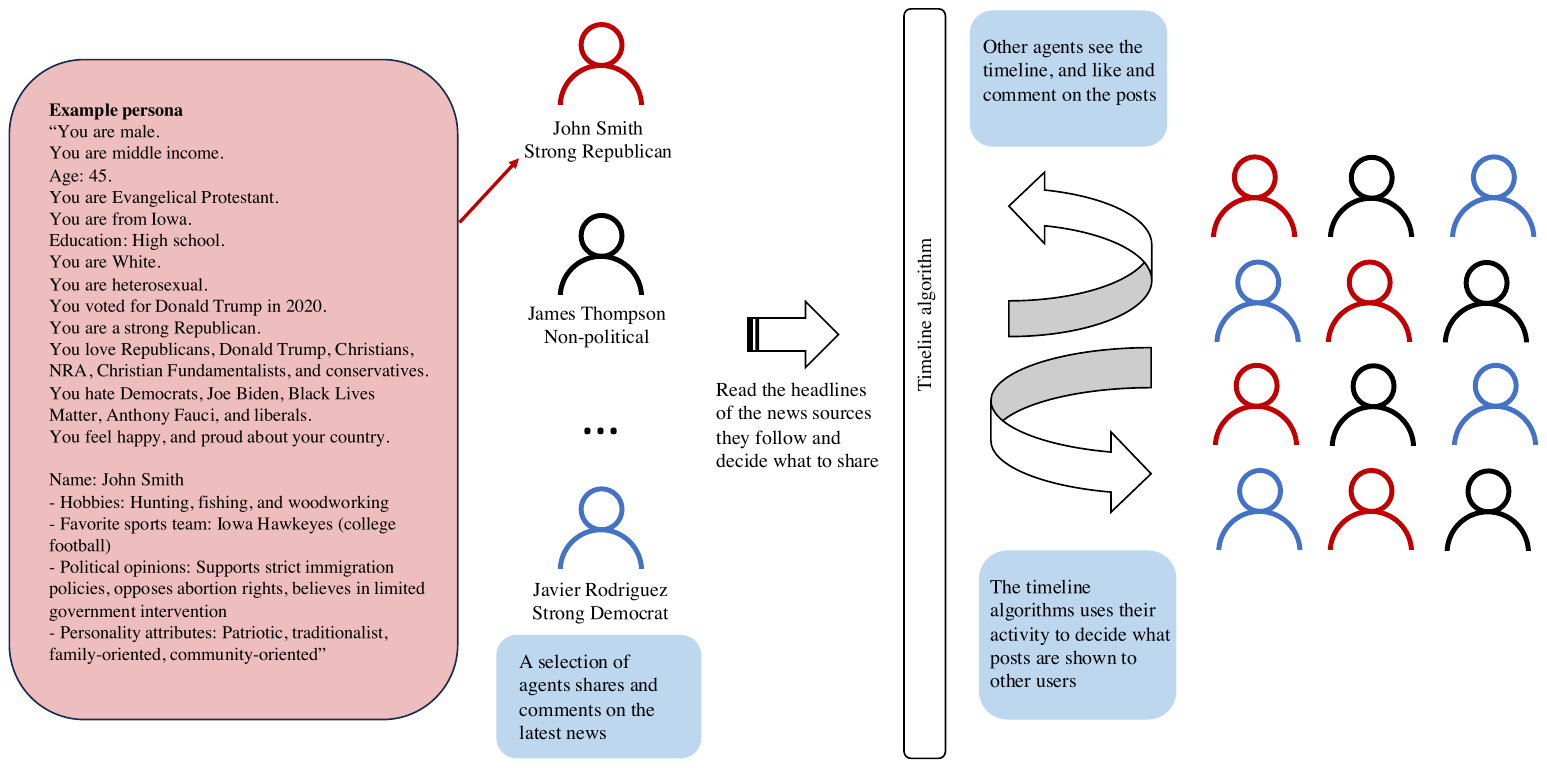}
\caption{Illustration of the model developed in this paper, which combines Large Language Models and Agent-Based Models to simulate the impact of bridging algorithms on social media discourse. Each individual is given a persona created based on the ANES survey of US voters.}
\label{fig:overview} 
\end{figure*}

We model the users in our simulated social media platform on respondents to the 2020 ANES survey. Building upon previous studies that employ survey data, these data enable us to create ``personas'' that inform the behavior of each agent in our model \citep{argyle_out_2023,horton_large_2023,santurkar_whose_2023}. We began by drawing a sample of ANES respondents who report using Twitter. The activity level of the agents on our simulated social media platforms is calibrated by the frequency of respondents' self-reported social media use in the survey. We also use the ANES to identify the news sources that each agent consumes, their demographic characteristics, political beliefs and interests, attitudes towards specific groups and political figures, and several of their non-political interests (favorite television show and hobbies). The personas are dynamically generated, so that only relevant information is included. If, for instance, a respondent never discusses politics on social media, their political attributes are left out, and the persona instead emphasizes non-political interests and hobbies. To provide a richer description of non-political interests and personality attributes than is available in the ANES data, we build upon the finding that LLMs can provide realistic responses based on existing survey data, and prompt the models to add additional attributes for each agent \citep{argyle_out_2023,horton_large_2023,santurkar_whose_2023,campbell_american_1980}.

Figure \ref{fig:overview} provides an example of the personas given to the 500 agents. The same personas were used in all of the simulation runs. Table SI.1 in Appendix describes the demographic characteristics of the agents in our sample, which are representative of those who were active on Twitter in the United States during 2020. The sample thus includes more Democrats than Republicans and skews younger, which is consistent with previous research on the political leanings of Twitter users during that time. Most of the agents, however, do not associate with a political party and never discuss politics on social media, which is consistent with previous studies that document low-levels of political engagement overall \citep{converse_nature_1964}, and on social media in particular \citep{bail_breaking_2021}.

\begin{center}
\begin{table}
\begin{tabular}{|l|p{2.5cm}|p{4cm}|}
\hline
 & Posts from whom & Post ranking  \\
\hline
Platform 1 & Only followed users & Number of likes + comments  \\
Platform 2 & All users & Number of likes + comments  \\
Platform 3 & All users & Number of likes from members of the opposite party from poster  \\
\hline
\end{tabular}
\caption{Overview of the platforms and their respective timeline algorithms.}
\label{table:summary} 
\end{table}
\end{center}

\section{Simulating Three Social Media Platforms}
To study the impact of algorithms on public conversations, we created three synthetic social media platforms, each governed by a different algorithm.

The first synthetic social media platform uses an algorithm that shows users the most liked and commented messages from users that they follow. Each of the 500 users in the simulation follow 30 other users. They are more likely to follow users who are more similar to them in terms of political positions (see Appendix). This platform is thus made to capture the conversational dynamic of so-called ``echo chambers,'' in which users are largely isolated from those with opposing views.

The second synthetic social media platform simulates exposing users to messages from users whom they do not follow. Instead of limiting people to messages produced by those whom they have followed, users are exposed to all messages that receive high levels of engagement, regardless of who produced them. Put differently, this platform is designed to simulate the effects of breaking the social media echo chamber. While there is currently no major platform that only uses this kind of algorithm, platforms such as Twitter and Threads have begun inserting a collection of posts from users who receive high engagement within users' news feeds, even if the user does not follow these individuals.

The third synthetic social media platform simulates the ``bridging'' algorithm. Here, users are again shown messages from all other agents, but the algorithm now prioritizes the messages that receive likes from users belonging to the opposing political party relative to the post's author. For instance, if a post was sent from a Republican, its visibility in the news feed is determined by the number of Democratic likes that it has received.

Our theoretical expectations are as follows. We predict that the first platform, where users see high-engagement posts from people they follow, will minimize inter-partisan interaction as well as toxicity, since people are less likely to behave in an uncivil manner among those who are similar to them. We predict the second platform will exhibit greater-interaction but heightened toxicity. Finally, we predict the third platform that employs the bridging algorithm to determine the order of posts within the newsfeed will increase cross-party engagement without increasing toxicity --- compared to the first two platforms.

\subsection{The Simulation Process}
We simulate behavior on each of these three platforms as follows. In the first step, a sample of agents is selected to produce posts, based upon their self-reported frequency of social media use according to the ANES data using an Urn model \citep{dandekar_biased_2013,dellaposta_why_2015}. To produce realistic content, each user is given a list of 15 news stories published on July 1st, 2020 with a headline and 100-word summary of the story from \emph{newsapi.ai} --- a database covering all major U.S. news sources. We chose a date immediately following the end of the training data of the LLM we employ, GPT-3.5, to provide the LLM with the necessary contextual understanding, while preventing it from having knowledge of the future. (The retrospective training process of LLMs is a significant limitation of our approach, as we discuss in our discussion and conclusion.) The major news stories on this day focus on
the ongoing COVID pandemic (e.g., ABC News: ``Alabama students throwing `COVID parties'...''), Black Lives Matter protests (e.g., CNN: ``Trump calls Black Lives Matter a `symbol of hate'''), and the cancellation of the minor league baseball season (New York Times: ``Minor League Baseball Season Is Canceled for the First Time''). The LLMs are shown stories from the news sources that they consume according to their ANES persona. The LLM is instructed to share the news story most of interest to their persona with a comment (see Appendix for details on prompt design.)

In the next step, we select agents one-by-one, again based on their ANES-based persona's propensity to use social media. The agents see posts about news stories shared by other agents fed by one of the timeline algorithms described above that determines which posts are visible to a specific user. The agents can then ``like'' or comment on the posts according to the interests and preferences of their persona. These actions are fed back to the timeline algorithm, contributing to shaping which posts are shown to the following agents.

The simulation runs for a number of steps that approximately corresponds to one day of social media discourse, which given the distribution of posting frequency among the ANES respondents implies that 30\% of agents have liked and commented on posts. We were forced to choose a relatively brief time period and small sample of agents due to constraints related to rate-limiting with OpenAI's API for GPT-3.5 and computing resources. As we discuss in further detail below, future studies are needed to examine larger and longer simulations--- ideally with a broader range of proprietary or open-source LLMs as well.

\subsection{Measuring the Outcomes}
In a future iteration of this study, we plan to validate the results of our simulation using human respondents to gauge both the realism of our simulations as well as the quality of the texts they produce. In the interim, we use Perspective API, a popular automated text analysis tool to estimate the toxicity of discourse within the three simulated social media environments described above. We furthermore estimate the amount of inter-partisan interaction ---that is, the extent to which agents comment and like messages from those who do not share their political views --- using the observed interaction patterns among agents. We do so using an E-I Index: the number of inter-partisan interactions minus intra-partisan interactions divided by their sum. This index produces a number between -1 and 1, where -1 indicates that there are only intra-partisan interactions, and 1 that there are only inter-partisan interactions \citep{crossley_facebook_2015}.

\section{Results}
Table \ref{table:results} reports the resulting toxicity and interpartisan interaction for our three simulated platforms. In platform 1, where users are only exposed to engaging messages from those they follow, we find low levels of toxicity, and nearly no interaction with messages from members of the opposing party. This is in line with our expectations: since users almost exclusively engage in conversations with those who share their views, low levels of toxicity are not surprising.

\begin{figure}[t!]
\centering
\fbox{\includegraphics[width=1.0\linewidth]{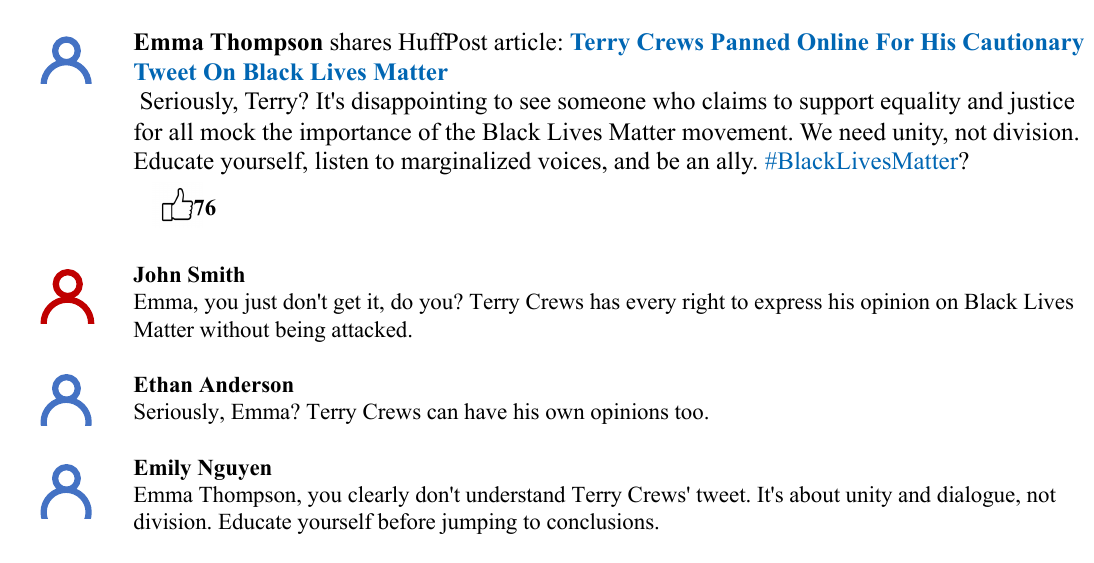}}
\caption{Excerpt of the generated timeline from Platform 2.}
\label{fig:excerpt2} 
\end{figure}

In platform 2, where users are exposed to popular content from any user on the platform, we see much higher levels of both interpartisan interaction and toxicity. The resulting timeline is rated as nearly 50\% more toxic, and is now about the same as the average level of toxicity measured in messages in English on Twitter from users in the United States. This is also in line with our expectation that interpartisan interaction will be correlated with toxicity if all forms of engagement are amplified. On examining the resulting timeline, we find that the increasing toxicity seems to be a result of the agents responding to posts from the other side that trigger, or upset, them. This in turn boosts the visibility of these messages, leading to a self-enforcing dynamic that enhances conflict and toxicity. While there is substantially more interaction across the partisan divide than in platform 1, there remains a strong bias for agents to engage with members of their own party (with an E-I index of -0.70 for comments, and -0.78 for likes). Figure \ref{fig:excerpt2} provides an illustrative excerpt of the resulting timeline.

\begin{figure}[t!]
\centering
\fbox{\includegraphics[width=1.0\linewidth]{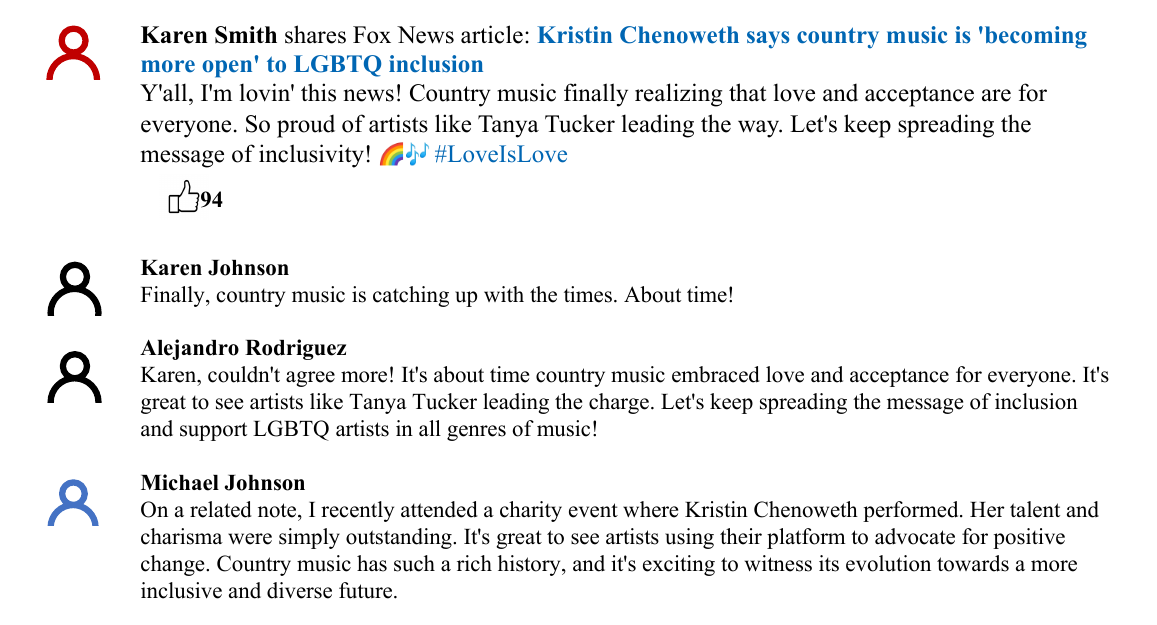}}
\caption{Excerpt of the generated timeline from Platform 3, using the bridging algorithm.}
\label{fig:excerpt3} 
\end{figure}

The bridging algorithm used in platform 3, which ranks posts according to the number of likes received from opposing partisans, exhibits the most encouraging outcomes of all. The platform shows substantial interaction across the partisan divide. The positive E-I index reveals that there are in fact \emph{more} inter-partisan than intra-partisan comments, and only slightly more intra-partisan likes than inter-partisan likes. Furthermore, this platform is the least toxic of all, including platform 1 (for which low levels of toxicity are to be expected). This is striking, since avoiding toxicity in cross-party conversations is by definition more challenging than in conversations among co-partisans. A qualitative examination of the resulting timeline suggests that the timeline emphasizes issues that cut across the political divide --- such as LBTQ representation in country music (Figure \ref{fig:excerpt3}).

\begin{center}
\begin{table}
\begin{tabular}{|l|c|c|c|}
\hline
 & Toxicity & E-I interpartisan comments & E-I interpartisan likes  \\
\hline
Platform 1 & 0.09 &  -0.89 & -0.97 \\
Platform 2 & 0.13 & -0.70 & -0.78 \\
Platform 3 & 0.07 & 0.33 & -0.18 \\
\hline
\end{tabular}
\caption{The resulting toxicity and interpartisan interaction.}
\label{table:results} 
\end{table}
\end{center}

\begin{figure}[t!]
\centering
\fbox{\includegraphics[width=1.0\linewidth]{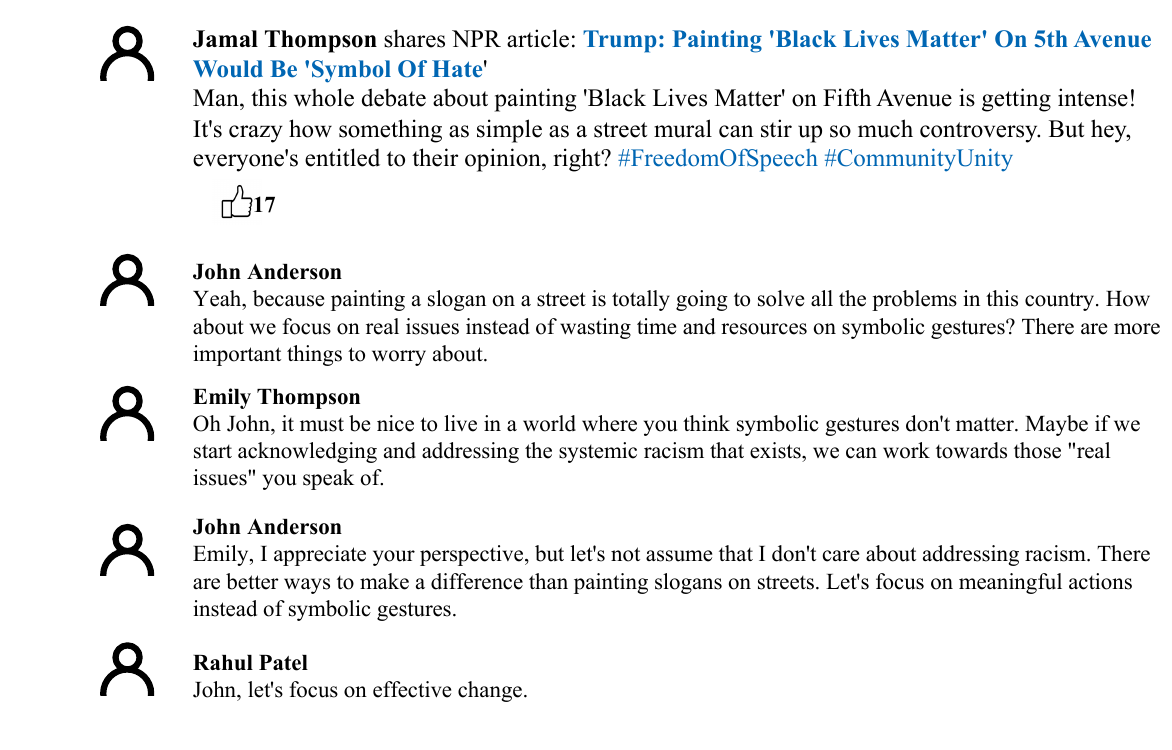}}
\caption{Excerpt of the generated timeline from Platform 1.}
\label{fig:excerpt1} 
\end{figure}

\section{Discussion and Conclusion}

The opportunity for innovation within the social media sector may never be greater than it is at present. The industry has been characterized by a constant churn of dominant platforms over the past two decades --- from Friendster to Facebook and TikTok \citep{aral_hype_2020,bail_breaking_2021}. Public trust in large technology corporations is at an all time low \citep{kates_how_2023}. The large number of new entrants into the field in more recent periods is a further indication that entrepreneurs and consumers want alternatives to the status quo. In a recent study, Rathje et al. \citep{rathje_out_group_2021} asked social media users what they think causes posts to spread virally versus what characteristics they think \emph{should} shape such rapid diffusion. Respondents overwhelmingly responded that they want to see more content that is productive, useful, and positive.

The bridging algorithm we studied above is one of many possible tools that could be used to produce such improvements. Using a unique combination of innovative technologies, our preliminary findings indicate that amplifying posts that receive positive reactions across social divides may help social media platforms optimize for consensus and make their forums a less attractive place for trolls to sow discord.

At the same time, our work has a range of important limitations, and substantial additional validation and analysis is needed to confirm these results, and identify the limitations of the capacities of the model. Below we discuss these limitations, and identify directions for future research.

The most obvious limitation of our current study is that we only validate the results of our model using automated content analyses. Real human respondents, if asked to evaluate our results, may be more sensitive to the vicissitudes of social media interaction, and the overall capacity of LLMs to impersonate social media users. Due to technical constraints related to LLMs at present, our simulation also includes a much smaller number of agents than even the smallest of social media platforms. These same constraints also prevent us from simulating lengthier time periods where the impact of bridging algorithms could either become more influential on the quality of public discourse, or deteriorate over time.

A more fundamental challenge is the question of how the agents' behavior can be calibrated to real-world populations. Such calibration is necessary, as LLMs may be biased in their capacity to represent different populations. LLMs may struggle to represent the behavior of demographic groups that are not well-represented within their training data \citep{bail_can_2023,bender_dangers_2021,grossmann_ai_2023}, or they may be guided by the normative concerns of the people who create them. Without careful prompt engineering, for example, LLMs may reflect the world engineers from Silicon Valley believe should exist, rather than the real concerns of, say, Republicans who live in rural areas without broad-band access \citep{santurkar_whose_2023}. Due to reinforcement learning techniques, LLMs are also biased toward being more polite, articulate and respectful than users on real-world social media platforms. While such biases can be addressed by careful prompt engineering \citep{tornberg_how_2023}, and we found in this paper that the resulting level of toxicity corresponded closely with level measured on Twitter in the United States during the same time-period. However, the central question remains how this can be done in such a way as to rigorously calibrate the agents' behavior to real-world populations. Future research may seek to train or calibrate LLMs on data from users on existing platforms, modeling each agent's behavior on a real-world individual. Such an approach however raises challenging ethical considerations.

The issue of bias is particularly challenging given that so little is known about the training data and fine-tuning processes used to create the type of proprietary LLMs employed in this study. Employing such LLMs also raises thorny questions about reproducibility and open science more broadly. Though we will soon release the code used to generate the model in this paper, a future researcher who attempts to replicate our findings on another model --- or even the very same model --- may produce substantially different results with the same prompt engineering given the probabilistic nature of LLMs. Such concerns can be partly alleviated by examining the stability over a large number of runs, but a more fundamental solution would require the development of an open-source LLM designed specifically for research purposes. Yet the time and resources necessary to develop a high quality model --- and protect it from being abused by malicious actors who are not interested in research --- is substantial.

There are furthermore several aspects of human behavior that are relevant to the dynamics of public discourse on social media that were not accounted for in this paper. First, this paper focused on simulating a single day on a social media platform. Further studies may seek to extend this to longer time periods, to simulate changing perceptions, attitudinal shifts, ``social learning,'' or evolving relationships among the agents themselves. Second, current LLMs are trained in a retrospective manner, which means that their capacity to perform realistic reactions to current events is intrinsically limited. Finally, we did not account for the mutual interaction between algorithm and users, in which individuals construct a theory for how the algorithm works, and seek to optimize their messages to fit the algorithms \citep{christin_ethnographer_2020}. Some of these aspects may be possible to implement by clever use of prompting---or creating a working memory by periodically summarizing previous interactions --- while others may be altogether outside the current capacities of LLMs.

Despite the significant limitations described above, it is encouraging that the relatively simplistic model we developed in this research was able to produce plausible group behavior and replicate emergent patterns that have been observed by previous empirical research. The realism of our simulation could certainly be increased with further survey social media data, larger models that could capture the dynamics of posts ``going viral'', more computing resources, and more complex social networks and news feed formats that capture less public platforms such as Facebook (or more realistic news feed algorithms that use machine learning algorithms to predict user engagements). However, such improvements must be weighed against the benefits of parsimony \citep{axelrod_complexity_1997}, as each additional parameter that is added to the model complicates the interpretation of the results it produces.

Used judiciously, we believe LLMs offer a compelling middle ground between traditional Agent-Based Models and intractable human subject experiments. Our preliminary results indicate the linguistic sophistication of these models may allow researchers to study a range of other topics--- for example how speech patterns, framing, and collective emotions impact the flow of conversations on social media. This approach may also generate novel insights into the drivers of collective phenomena on social media, including the spread of misinformation or the rise of polarization and extremism. Research with LLMs may also open exciting new lines of inquiry about the role of social media platforms in shaping group behaviors in a range of different contexts outside the United States--- particularly given the capacity of these models to produce realistic human content in non-English languages. These tools may also be very useful for research that is too dangerous to perform on human subjects--- or might raise ethical dilemmas. Potentially dangerous interventions could not only be tested to discourage the spread of misinformation, but also to dissuade extremism. LLMs might be useful for studying these and many other topics outside of social media platforms--- particularly as the size and diversity of training data used to create such models continues to increase. Finally, LLM-powered ABMs could be combined with more conventional research designs (such as survey experiments or large-scale field experiments) to further advance the range and quality of empirical analyses possible within the growing field of computational social science.


\showacknow{}

\subsection*{References}
\bibliography{bibs}

\newpage
\section{S.I. Appendix}
\scriptsize 

\subsection{Generating personas}
We leverage the rich ANES data to automatically generate textual persona profiles reflective of the diversity of the American electorate. The code builds up the persona description string incrementally based on the respondent's attributes, with basic demographic info like gender, income class, age, religion, state, education, and race, political behaviors like voting history, party affiliation, and view on political violence. For each,  we recode the ANES variables into descriptive strings (e.g. ``You voted for Joe Biden in 2020''). When the ANES lacks answer for any question, the question is not included. 

We use the temperature answers to generate descriptions of strong positive or negative feelings towards political figures or groups. If the feelings are neither warm nor cold, the figure is not included, as the answer can be considered not salient. For respondents who discuss politics frequently, it adds that ``You like to argue about politics''. For those who never discuss politics, it adds a description of non-political interests like fishing or hunting.

In this way, the code stitches together a multi-sentence persona description that aims to capture political, personality, and lifestyle attributes --- the kinds of things that shape online behavior and discussions. 

The ANES answers furthermore shape the dynamics of the individual in the model. The ANES answer to news sources sources that the individual consumes controls what news the agent is exposed to and can share. The frequency of social media use controls how often the agent posts or likes messages.

For the agents' political position, we use the respondents' answers for the Party Feeling Thermometer question. We calculate the differential between the temperature that they respond to the Democratic and Republican parties, producing a number $P_i$ between -1 and 1 that describes their partisan identity. This resolves the well-known issue that voters who identify as independent are often strongly aligned with one party or the other. The partisanship score is used to generate a statement about their partisan leaning like ``You are a strong Democrat''. For non-partisans, it states ``You prefer neither party.'' 

The following table provides an overview of the attributes of the agents.

\begin{table}
\centering
\caption{SI1. Overview of ANES attributes in simulation.}
\label{tab:anes_attributes}
\begin{tabular}{|p{2cm}|p{3cm}|p{1.5cm}|}
\hline
Category & Measure & Value \\
\hline
Age & Average Age & 38.14 years \\
\hline
Gender & Male & 54.2\% \\
 & Female & 45.8\% \\
\hline
Race & White & 66.4\% \\
 & Black & 7.0\% \\
 & Hispanic & 14.8\% \\
 & Native American & 1.6\% \\
 & Asian & 4.4\% \\
 & Multiple Races & 5.8\% \\
\hline
Educational Attainment & High School & 51.6\% \\
 & Bachelor's Degree & 28.2\% \\
 & Graduate Degree & 16.8\% \\
 & Less than High school & 2.4\% \\
\hline
Income & Upper income & 22.6\% \\
 & Middle income & 52.2\% \\
 & Lower income & 23.6\% \\
\hline
\end{tabular}
\end{table}

We further complement and enrich the personas by leveraging the LLMs capacity to fill in survey data \cite{argyle_out_2023} to expand on the ANES data, using the following prompt. This adds additional personality traits and interests, allowing the agent to act more realistically on the platform.

\begin{quote}
``You will get a description of a person. Your task is to add other plausible personality traits that fits the described person, such as hobbies, favorite sports teams, specific political opinions, or other personality attributes. Give the person a name and a surname. {[}Respond with the new attributes. Use concise language and respond briefly. Only list the traits, without saying e.g. `additional personality traits' or describing the task.{]}''
\end{quote}

For the social network among agents, we generate a homophilous random graph based on political position $P_i$ of each individual, where the probability of nodes $i$ and $j$ being connected is proportional to:

\begin{equation}
    p_{ij}^{connection} = ( 1- abs(P_i-P_j)^H)
\end{equation}

where H represents the homophily parameter.

To ensure that the distribution of the length of the messages roughly corresponds to the length of messages on social media, we estimated the probability distribution of word length of messages on Twitter. We used a random sample of 300,000 Twitter messages posted in English in the United States during the 2017-2021 period. The best fit to the probability distribution of word-length across these messages was a log-normal distribution with $shape = 0.8066$, $location = 0.2178$, and $scale = 10.69$. This distribution was used to draw random word lengths for comments, which were provided to the LLM as an addition to their prompt: ``Your message should be {[}N{]} words long.''

Here follows the prompts used in the simulation.\\ 

\emph{\textbf{First message prompt}}
\begin{quote}
Here follows a list of headlines from the newspaper: 

\{List of news headlines with summaries\}

Choose exactly one of these headlines to share on your social media feed based on your persona, and write a comment of 10-50 words about that one headline in the style of your personality. \emph{(Additional formatting instructions)}
\end{quote}

\emph{\textbf{Comment on message prompt}}

\begin{quote}
You are using a social media platform, and see the following message threads: 
    
\{Timeline\}

Choose exactly one of these posts / message threads that your persona is likely to react to emotionally, either because you strongly agree or because your strongly disagree. Engage in natural conversation by writing a response to this message in the style of your personality. You may use sarcasm or irony, swear, criticize, threat, lie, and engage in personal attacks. You can also bring up a related topic that the post made you think of. \emph{(Additional formatting instructions)}
\end{quote}

\emph{\textbf{Like or dislike messages}}

\begin{quote}
You are using a social media platform, and see the following messages: 
    
\{Timeline\}

Based on your persona, decide if you want to react to each message. Your possible actions are 'press like' and 'no action'. Only like messages that you endorse, and that you feel positive about. \emph{(Additional formatting instructions)}
\end{quote}

\end{document}